# Choice of Reference Surfaces to assess Plant Health through leaf scale temperature monitoring

**Rather Laasani Sanya Shabir, Shivani Chauhan, and Navneet Kumar**[*]

[1]Department of Mechanical Engineering, IIT Jammu, Jammu & Kashmir-181221, India

*correspondence to: navneet.kumar@iitjammu.ac.in

## ABSTRACT

We use infrared thermal imaging to remotely monitor the temperature of the leaves and plant which in turn is an indicative of their health and stress; in particular water stress. A series of experiments were conducted using Fluke TiX580 thermal imager on tomato plants to correlate the leaf temperature to different stages such as 'healthy', 'dying/wilted', and completely 'dead/dry'. The leaf temperature was compared to a series of paper-based reference surface temperature; these are of different colors and some are dry while some are wet. All the reference surfaces were insulated at the bottom ensuring the heat interaction between their top surfaces and ambient only. The surfaces were kept sufficiently far from one another. The healthy leaf temperature was found to be close to that of white dry and black wet reference surfaces whereas the wilted and dying leaves temperature were observed to range between yellow and red signifying that this temperature range can better predict the onset of water stress in the leaves. The completely dead/dry leaves were observed to range between green and blue dry surfaces, respectively. However, most of the dead leaf temperature data was found to be accumulated closer to the green surface signifying green dry surface can better indicate a dead leaf's condition. The dying leaves were observed to exhibit 8-10℃ higher temperatures as compared to the healthy leaves in similar ambient conditions. Temperature-based health assessment provides us with a timely intervention to prevent leaf death compared to the optical monitoring since IR images revealed elevated leaf's temperature 2-3 days before the optical unhealthiness (appearance of yellowness on the leaf) was noticed.

## 1. INTRODUCTION

The energy interaction between the plants and their surrounding environment takes place through two important phenomena, photosynthesis and transpiration through stomata. This transpiration, i.e., loss of water (which is supplied by the soil moisture) and gaseous exchange from leaves helps in regulating the canopy temperature required for their ideal metabolic activities via evaporative cooling. If the demand of water required for transpiration exceeds the available soil water content, the energy exchange equilibrium between them gets disturbed leading to eventually a decline in the transpiration rates and rise in temperature beyond a critical value (Gates 1964; Ehrler et al., 1978) causing internal water-temperature stress generation, thus water-stressed plants often exhibit higher leaf temperature. To know how much stress is being generated, researchers had distinctly investigated plant-atmosphere interaction and developed a variety of methods to monitor & evaluate stress in plants. For that, different techniques were adopted to monitor the crop health beginning from the use of thermocouples (Curtis, 1936; Ehrler, 1973; Pieters and Schurer, 1973) and leaf gas exchange measurements using contact sensors (Pereira et al.,1986) long before Infrared thermometers came into existence.

In this context, Penman (1948) and Monteith (1965) laid the foundational groundwork for understanding the environmental factors influencing evaporation rates from open water and bare soil, and Tanner (1963) related the plant temperature to the physiological processes. These studies provided the context for subsequent studies on plant thermography. Gates (1964) showed how temperature variations in plants can indicate their health and hydration status. Ehrler (1973) observed that the soil moisture content affects the leaf temperature and concluded that the temperature difference between a leaf and ambient decreases after irrigation, attains a minimum value after some days and increases again when the soil moisture is not sufficient to support the metabolic activities. They introduced a method for determining the infrared emittance of leaves which significantly contributed to the development of techniques and tools for measuring plant temperatures in agriculture. Priestley & Taylor (1972) focused on assessing surface heat flux and evaporation using large-scale parameters including solar irradiation. Since, evaporation from soils influence the plant water balance, Or et al. (2012) provided the concept of evaporative length which become a potential for providing simple and spatially extensive estimates of potential soil evaporative losses over large areas.

Idso et al. (1977, 1979, 1981) and Jackson et al. (1977, 1981) determined a Stress-Degree-Day index defined as $T_c - T_a$ (canopy temperature – air temperature) which was found to be influenced by environmental variables such as relative humidity, wind speed, net radiation and vapor pressure deficit. They utilized infrared thermometry to measure the canopy and foliage temperatures. In continuation to their previous investigations, they further established the quantitative correlation between plant thermography and crop harvests and developed a Crop Water Stress Index (CWSI)' (a parameter to depict the onset of stress) for three crops (Squash, Alfalfa & soya bean). A linear relationship between foliage-air temperature differential and VPD was observed for well-watered crops under all climatic conditions which lead to the existence of a criterion to identify healthy crops i.e. transpiring at their potential rate. The study also addresses normalizing the stress-degree-day parameter, considering the environmental variability. However, the empirical CWSI proposed by Idso was based on non-water-stressed baselines (for lower limit on canopy temperature) which changes with crop and season variation and determination of canopy's upper limit was itself ambiguous. To avoid these ambiguities and errors caused due to measurement in the aerodynamic parameters involved in the Jackson's CWSI, Qiu et al. (1996) developed a new simplified approach based on the canopy temperature, imitation leaf temperature, and air temperature. Additionally, these variations could also occur due to errors in the plants temperature measurements which is associated with leaf diffusive conductance also (Meyer et al., 1985; Idso et al.,1988; Jones, 1999). To reduces these errors, remote monitoring of crops through thermography (Idso et al., 1969; Inoue, 1986; Jones, 2004; Jones and Schofield, 2008; Jones et al., 2009; Costa et al., 2013) became pivotal contribution to the field of plant thermography leading to non-invasive plant health assessment. Based on the airborne radiometry and ground-based meteorological data, Jackson et al. (1987) evaluated evaporation from field crops.

All these canopy temperature measurements based studies under different water stress conditions suggested that the increment in water stress lead to increase in the canopy temperature directing to a relationship between plant water status and temperature and setting the stage for the application of aerial and non-destructive methods in agriculture (Ihuoma and Madramootoo, 2017) for continuous monitoring of crop health and to achieve better yield and quality. This non-invasive technique (IR thermography) become an important tool for investigating the thermal patterns of crops to identify the altered metabolic activity or localized changes in transpiration rates and hence stress-induced hotspots or cool area indicating localized stress.

Along with all these investigations, various reference surfaces (like soil, bright aluminium foil, white polystyrene foam, metallic surfaces etc.) were also investigated to get the wet and dry bulb temperature of the leaves under specific conditions required for estimating the CWSI. Table 1 highlights the notable studies with details of reference surfaces being used to calculate the stress index due to insufficient water/mineral supply. Recently, Narasegowda and Kumar (2021) proposed certain set of dry and wet reference surfaces

(white, yellow, blue, and black surfaced colours) having low thermal mass inertia, which provides the wet and bulb temperature quickly using infrared thermography.

Table 1 Notable studies with their brief details on reference surfaces used for different plants health status assessment.

| Author(s) | Reference surfaces used | | Temperature | Plants studied |
|---|---|---|---|---|
| | **Dry reference** | **Wet reference** | | |
| Idso *et al.* (1981) | Non-transpiring crop | Well-Watered crop | Infrared thermometer | Squash, Alfalfa & Soybean |
| Leinonen & Jones (2004) | Leaves covered with Vaseline (Petroleum jelly) | Leaves sprayed with water | Infrared Camera | Broad Bean (Vicia faba L.) |
| Costa *et al.* (2013) | • dry leaves<br>• dry cotton<br>• empty container<br>• dry tensiometer<br>• metal artificial leaves | • wet Leaves<br>• wet Cotton<br>• water filled container<br>• wet tensiometer<br>• artificial wet leaf | Infrared Camera | WT ecotype Columbia Arabidopsis thaliana rice. |

The present investigation utilized the IR thermography to assess plant water stress and to identify the appropriate acceptable surfaces to correlate the leaf temperature at different stages such as 'healthy', 'dying/wilted' and completely 'dead/dry'.

## 2. MATERIALS AND METHODS

We used fresh Tomato plants for investigating the temperature characteristics of the leaves and relate it with the extent of water stress. Fluke TiX580 thermal camera (sensitivity of 0.05K and a pixel resolution of 640 x 480) was utilized to measure the temperature of the leaves and other surfaces of our interests. The ambient conditions like temperature, velocity, etc. were also monitored continuously. Though, the previous studies suggest that leaf health status is greatly affected by the moisture availability in the soil but we didn't monitor the moisture content for the present investigation since the objective of this study was restricted to find out the most suitable reference surfaces to indicate the health status of a leaf. Further, all the acquired infrared images underwent rigorous analysis using the Fluke Connect software platform for which emissivity adjustments are a crucial aspect. In this study, the emissivity value for the surfaces (Dry & Wet) is taken to be 0.97 and for plants the value is 0.94 (Woolley, 1971). The detailed procedure is seen in the supplementary material.

The paper-based artificial surfaces used in this study consists of a variety of 'dry' and 'wet' surfaces (see Fig.1) that were suitably mounted on a portable platform. These surfaces have low thermal mass and that make them highly sensitive to the atmospheric changes. The set of "dry" reference surfaces consist of six distinct colors, namely black, yellow, white, red, green, and blue, with each material unit characterized by dimensions measuring 1.1cm x 1.1cm x 2cm. These reference surfaces were deliberately furnished with an underlay of thermal insulation on all four sides to avoid thermal exchange with the ambient environment and systematically mounted onto a wooden substrate enveloped in reflective aluminum foil. These surfaces, thus, only have their top colored side exposed to the ambient. A separation of ~5cm was provided between the adjacent surfaces to effectively isolate them thermally similar to Narasegowda and Kumar (2021).

Concurrently, a set of uninsulated surfaces, doubly exposed, were affixed onto miniature wooden fixtures, each measuring 5cm in length and 1.2cm in width, cloaked in a reflective covering. These fixtures were arrayed along a 60cm linear wooden scale, uniformly spaced at 5cm intervals. In parallel, a comprehensive temperature assessment was conducted on five diverse shades of green (see Fig. 1g and Fig. 1h). Dark Green shade was selected for further study since the difference in temperatures of these surfaces were negligible. Furthermore, eight distinct "wet" reference surfaces were designed and classified into three distinct configurations. The initial configuration was designed to emulate a water surface akin to a vast water body, comprising a black-painted, water-filled, thermally insulated Petri dish with a volumetric capacity of 48.6cm$^3$. This configuration was supplemented with three distinct small bottle lids (each with a volume of 4.2cm$^3$) featuring colors of red, blue, and white; all devoid of thermal insulation. The second configuration aimed to replicate the water droplets adhered to a leaf's surface comprising discrete water droplets on a black paper substrate measuring 6cm x 4.5cm. The final configuration encompassed porous substrates, fully wet with water. These substrates were white, green, and coffee filter paper (adhered onto a cardboard enshrouded in reflective aluminum foil) with each having dimensions of 6cm x4.5cm. The brief description of all the surfaces used in the study is given in Table 2.

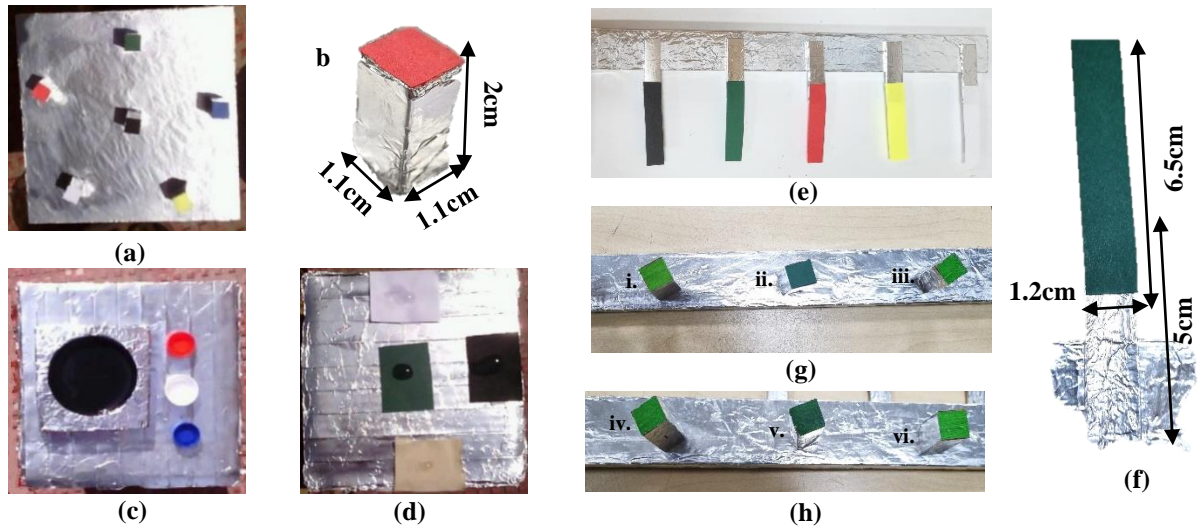

Figure 1 Various 'dry' and 'wet' surfaces used in the current investigation (a) shows all the thermally isolated dry surfaces systematically pasted on an insulating wooden fixture, starting with white, yellow, blue, green, red, and black in the centre, (b) all sides (1.1cm×1.1cm×2cm) of these surfaces covered with aluminium foil except the top coloured surface to prevent unnecessary thermal exchange, (c) coloured surfaces filled with water (d) four (6.5cm*4.5cm) different colours with water droplets sitting on them mimicking the droplets on leaf surfaces, (e) equidistant doubly exposed surfaces (6.5cm*1.2cm) on a wooden fixture, (5cm long), (f) clearly shows the dimensional feature of a doubly exposed surface, (g) and (h) six different shades of green colour, out of which the dark green shade marked as (v) was selected for further study.

The wet paper surfaces were thoroughly saturated with water, and a deliberate waiting period of 1 to 2 minutes was given to ensure the steady state. This preliminary step was essential for accurate temperature measurements. The underlying rationale for this protocol was to allow the wet surfaces to stabilize, attaining a consistent and predictable temperature. To fulfil this purpose, the thermal behaviour of the wet surfaces was closely monitored for approximately 10 minutes, commencing immediately after they were wetted. The progression of temperature changes was documented in a video recording, which was subsequently subjected to analysis using the Fluke Connect software. The temperature versus time curve shows the steady state time after 2.1 minutes (see Fig. S1 in the Supplementary material).

Two tomato plants housed in pots with a diameter of 170cm were used in our experiments. These plants underwent thorough monitoring between from April and June months of the year 2023. These experiments were intended to allow the plants to die. For this purpose, the experiments were conducted in three different stages as discussed below:

*i. Growth Stage*

In this stage, the plants were regularly watered to ensure their healthiness. The temperature data of the leaves were regularly monitored post acquiring them, i.e., after 4 April 2023. The surface temperature data of all the "wet" reference surfaces were also measured regularly to discern the optimal wet reference surface. The water-filled containers were exposed to the sun for a period of 5-7 minutes prior to temperature measurements. We observed that the temperature fluctuations in the water containers remained within a marginal range of $0.05^0C$ to $1.5^0C$. Furthermore, a comparison of temperature differentials among the various containers demonstrated that the black-insulated Petri dish, uninsulated black bottle lid, and black had negligible temperature variations in similar ambient conditions. In contrast, a substantial temperature difference was observed among the remaining surfaces, with the white wet surface exhibiting the least temperature values. Accordingly, the water-filled containers were excluded from further analysis. For about 5-6 days, even after stopping the water supply, (Fig. S2 in the supplementary section) the leaves remained healthy (as indicated by their temperature values); more discussion on this matter soon.

Table 2 Details of various dry and wet surfaces considered for the present investigation. All these surfaces have low thermal mass and can be considered as lumped mass. In general, these surfaces respond quickly to the atmospheric changes.

| Surface | Type | Colour | Dimensions | Details |
|---|---|---|---|---|
| Dry References | Insulated Cuboid with top square surface | White | Length = 1.1cm Breadth = 1.1cm Height = 2cm | Made from papers Insulated at their bottom Kept 5cm apart on wooden board wrapped in aluminium foil |
| | | Yellow | | |
| | | Red | | |
| | | Green | | |
| | | Blue | | |
| | | Black | | |
| | Doubly Exposed Rectangular surfaces | White | Width = 1.2cm Length = 5cm | Each paper strip, pasted on wooden scale wrapped in aluminium sheet |
| | | Yellow | | |
| | | Red | | |
| | | Green | | |
| | | Black | | |
| Wet References | Cylindrical Containers | Black | Diameter = 2.5cm thickness = 1mm | Small uninsulated bottle lids of volume 4.2cm$^3$, filled with water |
| | | Red | | |
| | | Blue | | |
| | | White | | |
| | | Insulated Petri Dish (Black) | Diameter = 7.9cm Thickness = 2mm | Water filled, insulated Petri dish of volume 48.6cm$^3$, Painted Black |
| | Rectangular Surfaces | White | Length = 6cm Breadth = 4.5cm | Paper sheets, pasted on cardboard wrapped in aluminium foil, wetted with water |
| | | Coffee FP | | |
| | | Green | | |
| | | Black | | |

*ii. Dying Period*

Around June 2nd, the plant leaves started to change their color showing the symptoms of water deficit and leading to water stress generation within them. However, the thermal images revealed that the surface temperature started to increase 2-3 days before showing the evidence of stress generation optically. At this point, if watered, they regain their health status. At this stage, dry reference surfaces were also brought into picture along with the wet reference surfaces to correlate the maximum and minimum attainable temperature by the leaves at this stage.

*iii.* *Death Period*

Once the leaves start to change their color and if they are not watered within certain time frame, the stress reaches to a critical value from where they would not regain their health (Parkash and Singh, 2020) and are considered to be dead. In our experiments, this stage happened after 4-5 days as evident by their surface temperature (refer Fig. 14). The temperature acquisition of wet and dry reference surfaces was also continued till the complete plant become dead.

## 2.1 Average Temperature: a suitable indicator of surface temperature.

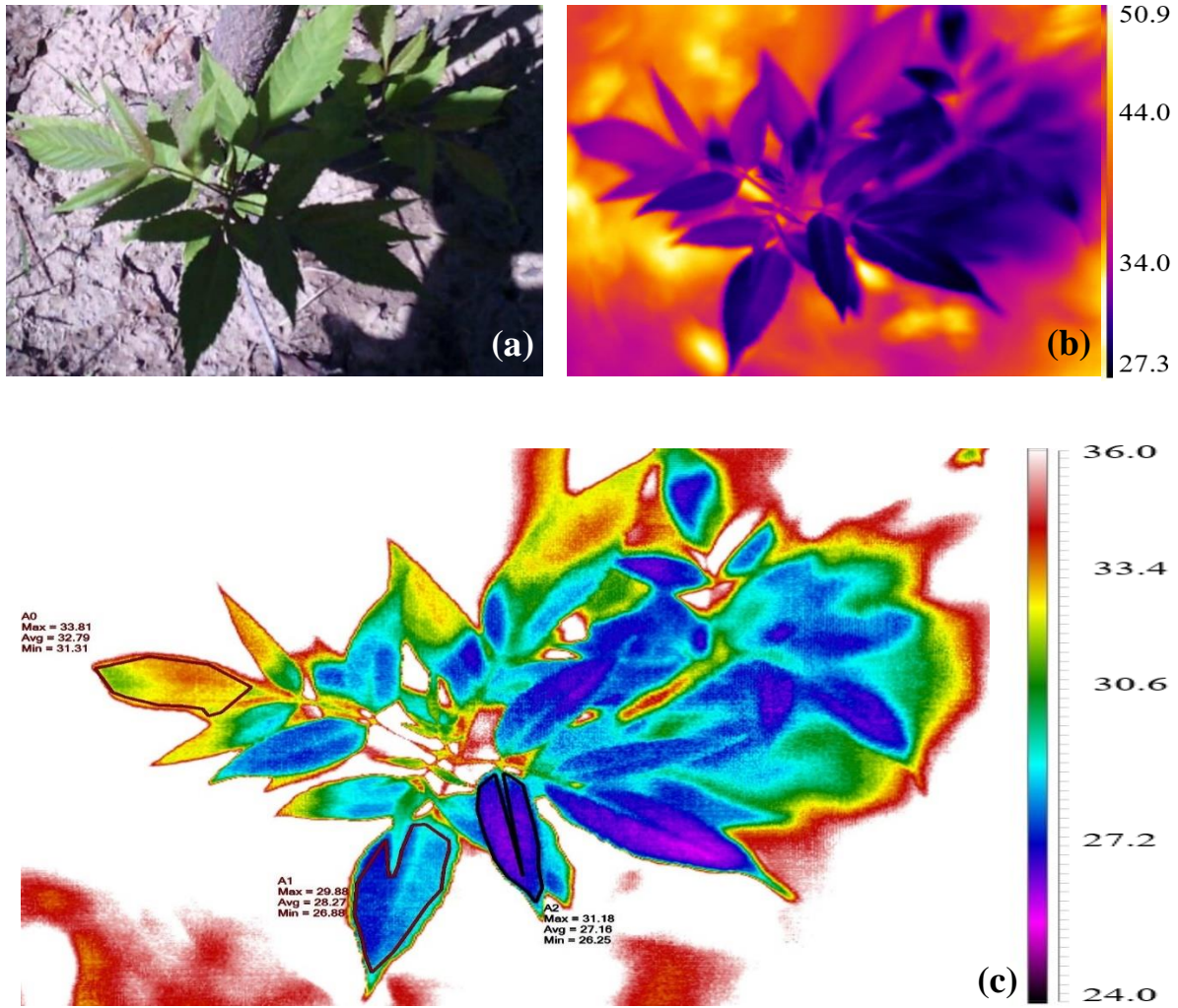

Figure 2 (a) Optical image of plant, (b) a raw image captured by the thermal camera in sunlight, and (c) the post-processed image showing the temperature contours along with a temperature scale. Here, the edges of the leaves were detected manually as seen from the three sample leaves. Also shown are the corresponding minimum, maximum, and average temperature values for reference. Note the temperature scales besides the thermal images that indicate 'magenta' and 'red' as the 'cold' and 'hot' spots, respectively.

The average temperature is a simple and quick way to understand the overall thermal behaviour of a surface (including the plant as seen in Fig. 2). Fig. 2a and 2b shows the optical and thermal images respectively, captured at the same instant and Fig. 2c shows the post-processed thermal image. The temperature scale, in Fig. 2c, ranges between 24⁰C and 36⁰C. We mark three different leaves, namely A0, A1, and A2. The temperature ranges in the same order are: $32.55 \pm 1.25^0 C, 28.38 \pm 1.5^0 C,$ and $28.73 \pm 2.5^0 C$ and the average temperature values are $32.79^0 C, 28.29^0 C$ and $27.16^0 C$ respectively. It is clear that a temperature range of $\pm 2.5^0 C$ is sufficient to capture the temperature variations of a leaf in the present investigation. Further, the majority of the leaves seen to lie within $\pm 1.5^0 C$ (see Fig. 2c).

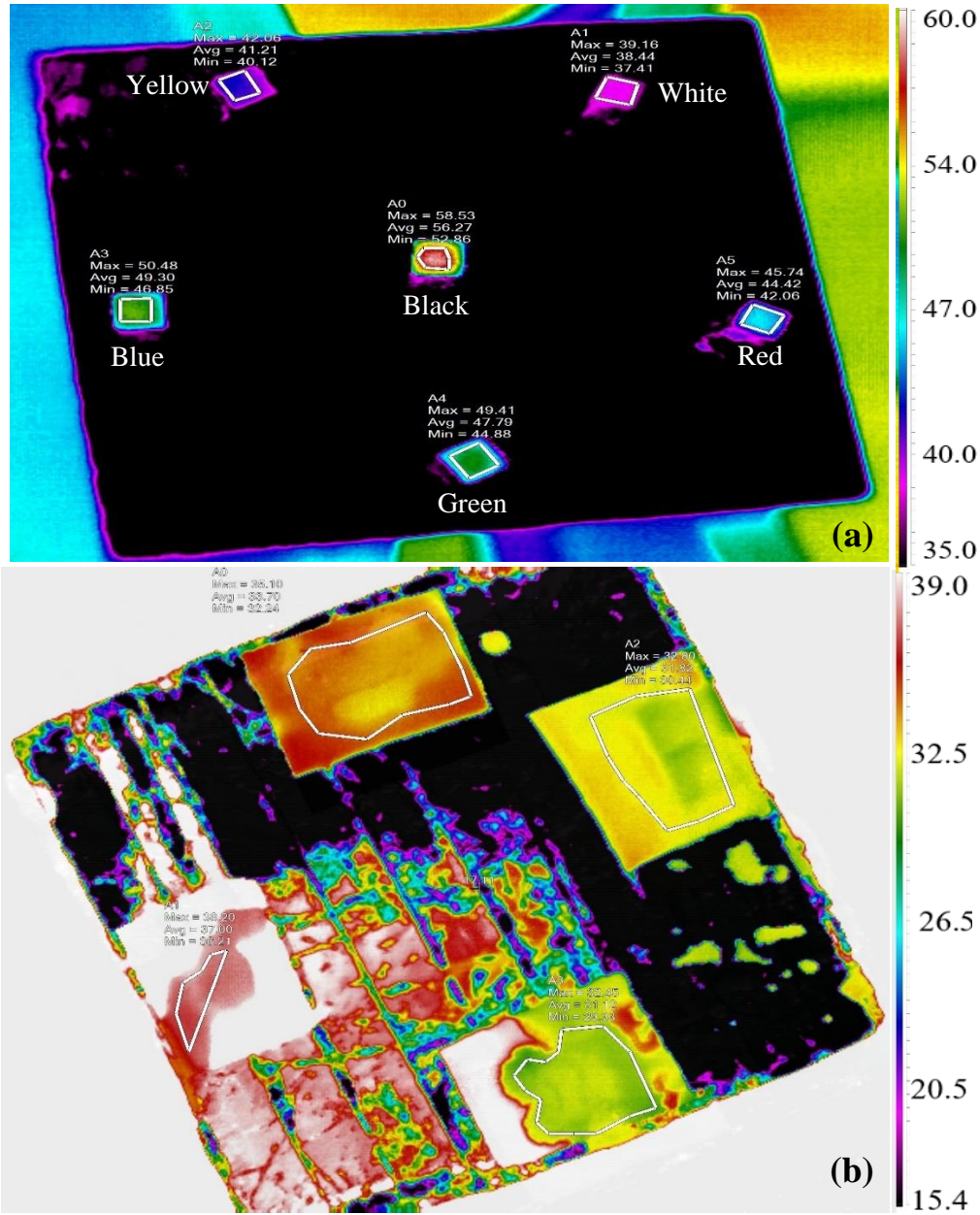

Figure 3 (a) and (b) show the IR images of 'Dry' and 'Wet' surfaces respectively. Small white boxes inside the different dry surfaces represent the region selected for obtaining the average temperature value. These boxes were 10×10 pixels, in size and 1.1cm×1.1cm in actual size. The wet surfaces are quite large in size and therefore only a relevant section was chosen to indicate the average temperature values. Note that, with time, the wet surfaces become dry and care must be taken while using them as reference surfaces.

The single average surface temperature value simplifies the complexity of thermal data into a straightforward metric and proves valuable for tracking and monitoring temperature variations trends followed by dying plants. Thus, comparing average temperatures over time or between different locations will provide insights into changes in thermal behaviour of plants and reference surfaces.

Figure 3 represents the results of post-processing of thermal image of all the dry surfaces and wet surfaces. The temperature data including the maximum, minimum and average values over the selected area of each surface is shown beside each surface. The difference in the maximum and minimum temperatures computed over specific area for each surface i.e. leaf, 'dry' & 'wet' surface and is defined as $\Delta T = T_{max} - T_{min}$ ($^0$C). This parameter is plotted against the average temperature for all three categories of surfaces mentioned above.

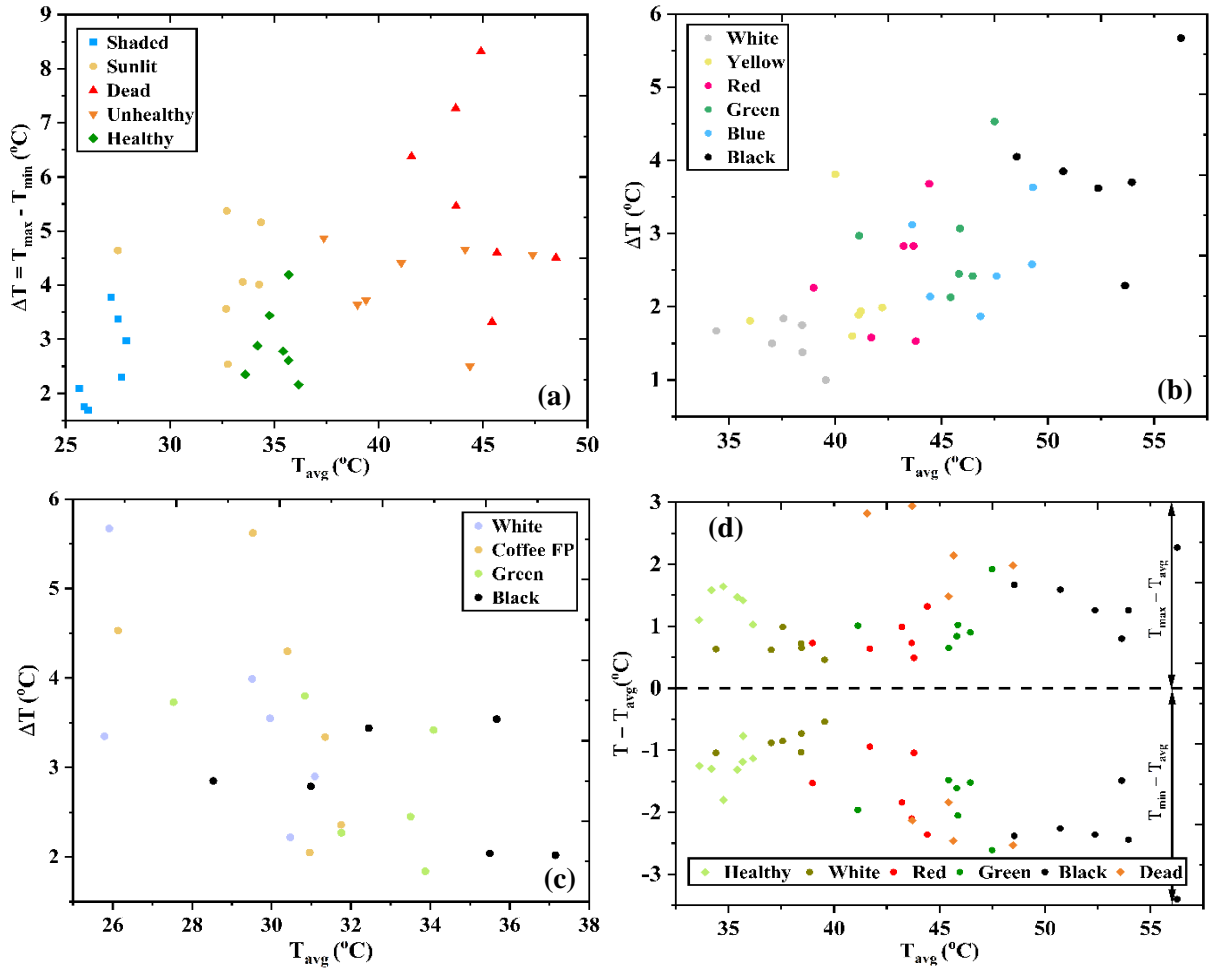

Figure 4 (a) Maximum and minimum temperature difference exhibited by the leaves in different conditions, (b) the corresponding temperature difference in dry reference surfaces, (c) temperature difference in wet reference surfaces, (d) temperature difference between the maximum/minimum and the average value of a few selected surfaces. The x-axis in this figure is the average temperature values.

The healthy leaves had low $\Delta T$ values (between $2^0$C and $4^0$C) and the sunlit leaves had slightly higher $T_{avg}$ compared to the shaded ones but had slightly higher $T_{avg}$ compared to the shaded ones but had similar $\Delta T$ range (see Fig. 4a). The higher temperature differences are exhibited by the dead leaves . Figure 4b & 4c show $\Delta T$ Vs Average temperature plot for all the 'dry' and 'wet' surfaces. It can be seen that most of the surfaces fall within a range of 1.6 - $4^0$C. Figure 4d represents the deviation of maximum and minimum temperatures from the average values for particular surfaces. It was observed that the surfaces with low average temperatures (e.g. healthy leaves and white dry surface) exhibited less deviations and had a narrow temperature range as compared to the surfaces having higher average temperature as in case of black dry surface and dead leaves.

## 3. RESULTS & DISCUSSIONS

### 3.1 Thermal response of various reference and leaf surfaces

Thermal images of tomato plants along with the dry surfaces were captured beginning from the month of April. Later, the wet surfaces were introduced to the experimental setup in the month of May & June, to subsequently measure the lower temperature limit in the plant vicinity. All the experiments were conducted within the time frame 11:00am -3:00pm. This effort was directed towards identifying the optimal candidates (reference surfaces) that could effectively serve as benchmarks for gauging the health status of a plant. The table below represents the dates on which experiments were conducted and the ambient conditions, along with the maximum and minimum plant temperature recorded on the respective day. The thermal examinations were conducted on the tomato plant at various stages and in order to record the maximum temperatures attained by plant leaves till it approached death. Table 3 shows the temperature record of the leaves under study during their growth period until the plant was kept well-watered.

Table 3: Experiment data record throughout the investigation period, viz. 6$^{th}$ April to 27$^{th}$ June, 2023. Also seen are the corresponding ambient temperature values.

| S. No | Date | Plant Temperature ($^0$C) | | $T_{amb.}$ ($^0$C) |
|---|---|---|---|---|
| | | $T_{max}$ | $T_{min}$ | |
| 1 | 6th April | 41.24 | 33.87 | 24 |
| 2 | 10th April | 41.66 | 36.37 | 28 |
| 3 | 12th April | 44.4 | 38.18 | 30 |
| 4 | 14th April | 43.36 | 36.61 | 30 |
| 5 | 5th May | 42.62 | 33.69 | 28 |
| 6 | 6th May | 39.86 | 31.69 | 29 |
| 7 | 8th May | 32.26 | 29.33 | 29,windy |
| 8 | 9th May | 44.49 | 29.67 | 28 |
| 9 | 15th May | 42.04 | 34.55 | 36 |
| 10 | 26th May | 37.26 | 30.27 | 27 |
| 11 | 27th May | 42.12 | 33.29 | 32 |
| 12 | 3rd June | 44.9 | 32.79 | 31 |
| 13 | 5th June | 49.28 | 41.08 | 34 |
| 14 | 8th June | 48.49 | 43.69 | 37 |
| 17 | 27th June | 46.33 | 42.13 | 34 |

#### 3.1.1 *Healthy leaf* versus dry and wet reference surfaces

Figure 5a shows the optical image of the healthy tomato plant, (Fig. 5b and 5c are its raw and processed thermal images captured in sunny conditions). Note, that in Fig. 5c, all the leaves (13 in numbers) were tracked in the software using the polygon feature. The minimum, average, and maximum temperature values of all the leaves are also seen. Also, the dry surfaces on the portable device were kept near the plants but sufficiently far enough from the walls and ground in order to avoid thermal interferences from the surroundings. Table 4 shows temperature data of all the 'dry' and 'wet' surfaces along with that of the healthy leaves. As clearly seen in Fig. 5d and Table 4, the minimum temperature (yet higher than that of ambient) was exhibited by the white dry surface followed by yellow and red surfaces, whereas, the maximum temperature was attained by the black dry surface. The healthy leaves were found to range between the wet black surfaces temperature and white dry surfaces temperature, however on extremely hotter days the healthy leaves sometimes attained 1-2$^{0}$C higher temperatures than the white dry surfaces.

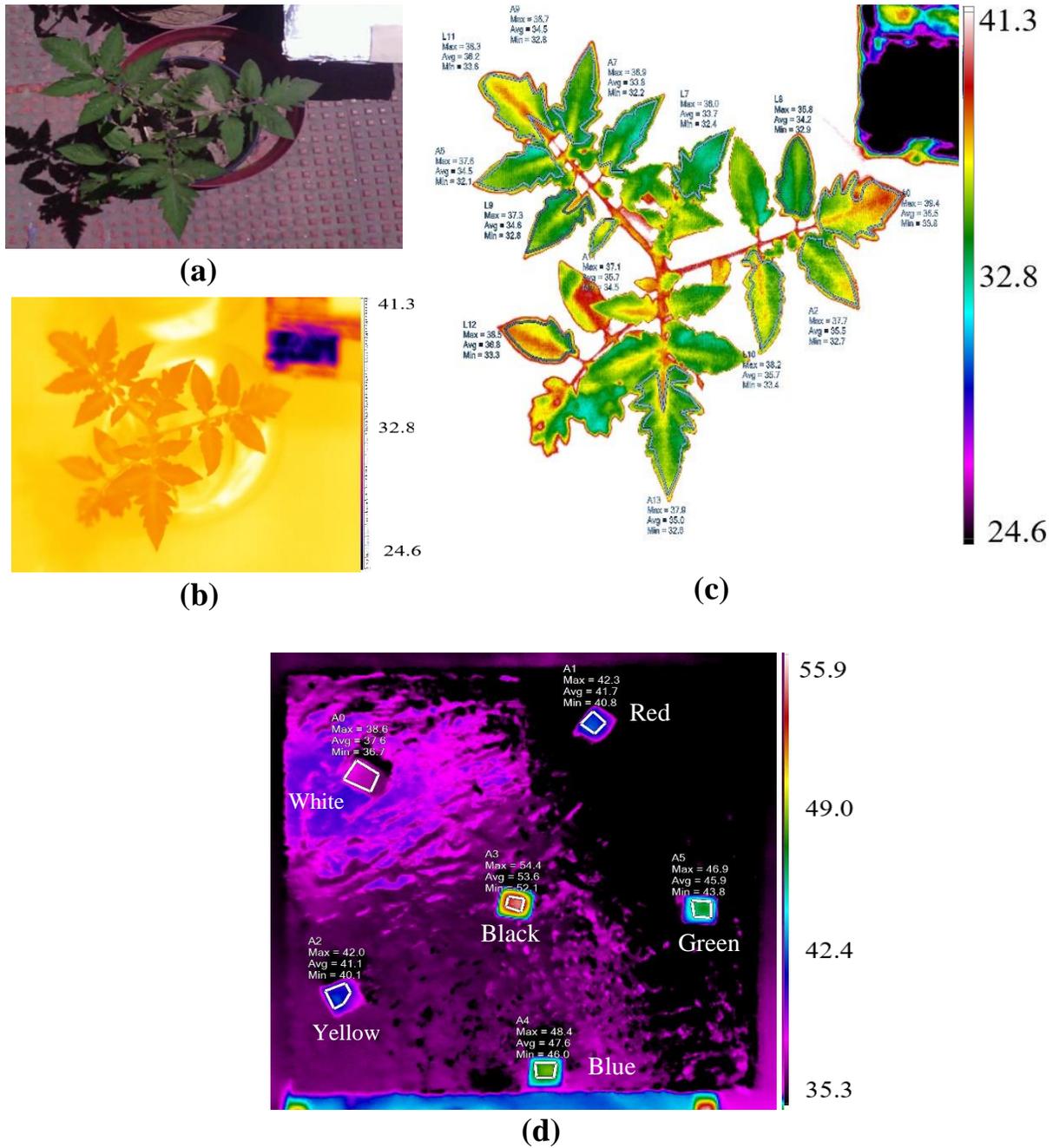

Figure 5 Comparison of the 'healthy' plant leaf temperatures' with a set of dry surfaces. (a) Visible image of plant under study, (b) raw thermal image of the plant captured on 5$^{th}$ May 2023, (c) post-processed thermal image of the plant showing individual leaves along with their minimum, average and maximum temperatures (here the boundary of the leaves within the thermal window were tracked manually using the polygon feature in the software), and (d) processed thermal image of all the dry surfaces. Also seen are the temperature scales besides all the thermal images.

Table 4: Surface temperature of *healthy* tomato plant leaves along with different thermally isolated 'Dry' & 'Wet' reference surfaces which were kept near the plant, ensuring similar natural conditions.

| Temperature °C (leaf) | Temperature °C (dry surfaces) | | | | | | Temperature °C (wet surfaces) | | | |
|---|---|---|---|---|---|---|---|---|---|---|
| **Healthy Leaf** | **White** | **Yellow** | **Red** | **Green** | **Blue** | **Black** | **White** | **Coffee FP** | **Green** | **Black** |
| 32.79 | 34.41 | 35.99 | 38.98 | 41.13 | 43.62 | 48.54 | 25.91 | 26.13 | 27.53 | 28.53 |
| 33.61 | 34.41 | 35.99 | 38.98 | 41.13 | 43.62 | 48.54 | 25.91 | 26.13 | 27.53 | 28.53 |
| 34.09 | 34.41 | 35.99 | 38.98 | 41.13 | 43.62 | 48.54 | 25.91 | 26.13 | 27.53 | 28.53 |
| 35.68 | 37.57 | 41.1 | 41.7 | 45.87 | 47.57 | 53.63 | 30.47 | 30.96 | 31.76 | 32.45 |
| 35.69 | 37.03 | 40.01 | 43.68 | 45.82 | 44.46 | 52.37 | 25.79 | 29.53 | 30.84 | 30.99 |
| 35.7 | 37.57 | 41.1 | 41.7 | 45.87 | 47.57 | 53.63 | 30.47 | 30.96 | 31.76 | 32.45 |
| 35.9 | 37.03 | 40.01 | 43.68 | 45.82 | 44.46 | 52.37 | 25.79 | 29.53 | 30.84 | 30.99 |
| 36.01 | 38.46 | 42.21 | 43.22 | 46.46 | 49.25 | 53.95 | 29.51 | 30.4 | 34.08 | 35.5 |
| 36.16 | 38.46 | 42.21 | 43.22 | 46.46 | 49.25 | 53.95 | 29.51 | 30.4 | 34.08 | 35.5 |
| 36.42 | 38.46 | 42.21 | 43.22 | 46.46 | 49.25 | 53.95 | 29.51 | 30.4 | 34.08 | 35.5 |
| 36.66 | 37.03 | 40.01 | 43.68 | 45.82 | 44.46 | 52.37 | 25.79 | 29.53 | 30.84 | 30.99 |
| 36.76 | 37.57 | 41.1 | 41.7 | 45.87 | 47.57 | 53.63 | 30.47 | 30.96 | 31.76 | 32.45 |
| 36.85 | 37.03 | 40.01 | 43.68 | 45.82 | 44.46 | 52.37 | 25.79 | 29.53 | 30.84 | 30.99 |
| 36.96 | 38.46 | 42.21 | 43.22 | 46.46 | 49.25 | 53.95 | 29.51 | 30.4 | 34.08 | 35.5 |
| 37.54 | 37.03 | 40.01 | 43.68 | 45.82 | 44.46 | 52.37 | 25.79 | 29.53 | 30.84 | 30.99 |

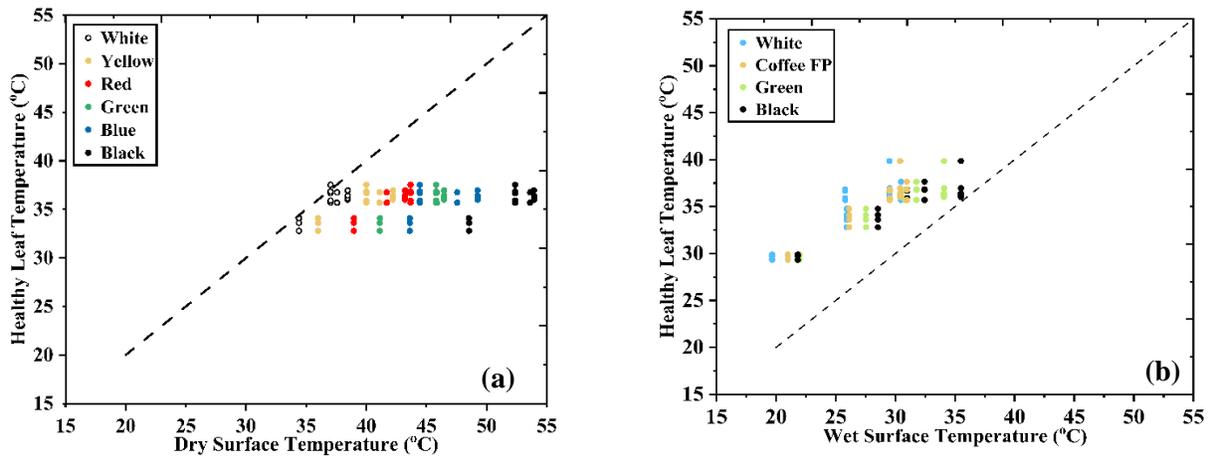

Figure 6 Comparison of *healthy leaves* temperature data with (a) dry and (b) wet reference surfaces using $45^0$ slope line. The temperature of the healthy leaf seems to be very close to 'white dry' and 'black wet' surfaces. As expected, the 'black dry' and 'white wet' surfaces attain the highest and lowest temperatures respectively.

Now, we compare the healthy leaf temperature values with those of the dry and wet reference surfaces. For this purpose, we plot healthy leaf temperature versus all the dry surfaces (Fig. 6a) and all the wet surfaces (Fig. 6b) used in this study along with the corresponding $45^0$ lines. The point of coinciding with this slope line would mean equal temperature of that particular surface and the healthy leaf. It can be seen that healthy leaf temperature values are mostly left and right side of the slope line when compared with the dry and wet reference surface temperature values, respectively. The maximum deviation from the slope line is seen in the case of black dry surface and at the same time, white dry surface shows the minimum variation from the healthy leaf surface (see Fig. 6a). Similarly, it is evident from Fig. 6b that the minimum temperatures attained by the healthy leaves on a particular day is closer to that of the black wet surface. However, the lowest temperatures were observed to be that of the white wet surface, i.e. 6-7°C below the minimum healthy leaf temperature. We here conclude that the temperature values of the healthy leaves are close to those achieved by 'white dry' and 'black wet' reference surfaces.

### 3.1.2 *Dying leaf* versus dry and wet reference surfaces

It has been observed in the previous section that the white dry' and 'black wet' reference surfaces temperature agrees reasonably well with that of the healthy leaf temperatures. Further, to find the surfaces that would indicate the dying or dead leaves, we intentionally stopped watering the plant after they were fully grown, inducing water deficit in them. Figure 7 shows the optical and IR images of plant and dry reference surfaces on a particular day as the plant started dying and exhibited higher temperatures. Some leaves can be seen to have become yellowish in the optical image (Fig. 7a), the similar leaves can be seen to exhibit higher temperatures in the processed thermal image (Fig. 7c). The range of the temperature scale can be clearly seen to have shifted towards higher temperatures; most of the leaves, even the ones that appear green in the visible image have temperatures in the range 40-44$^0$C, i.e., almost 8-10$^0$C higher than the healthy leaves.

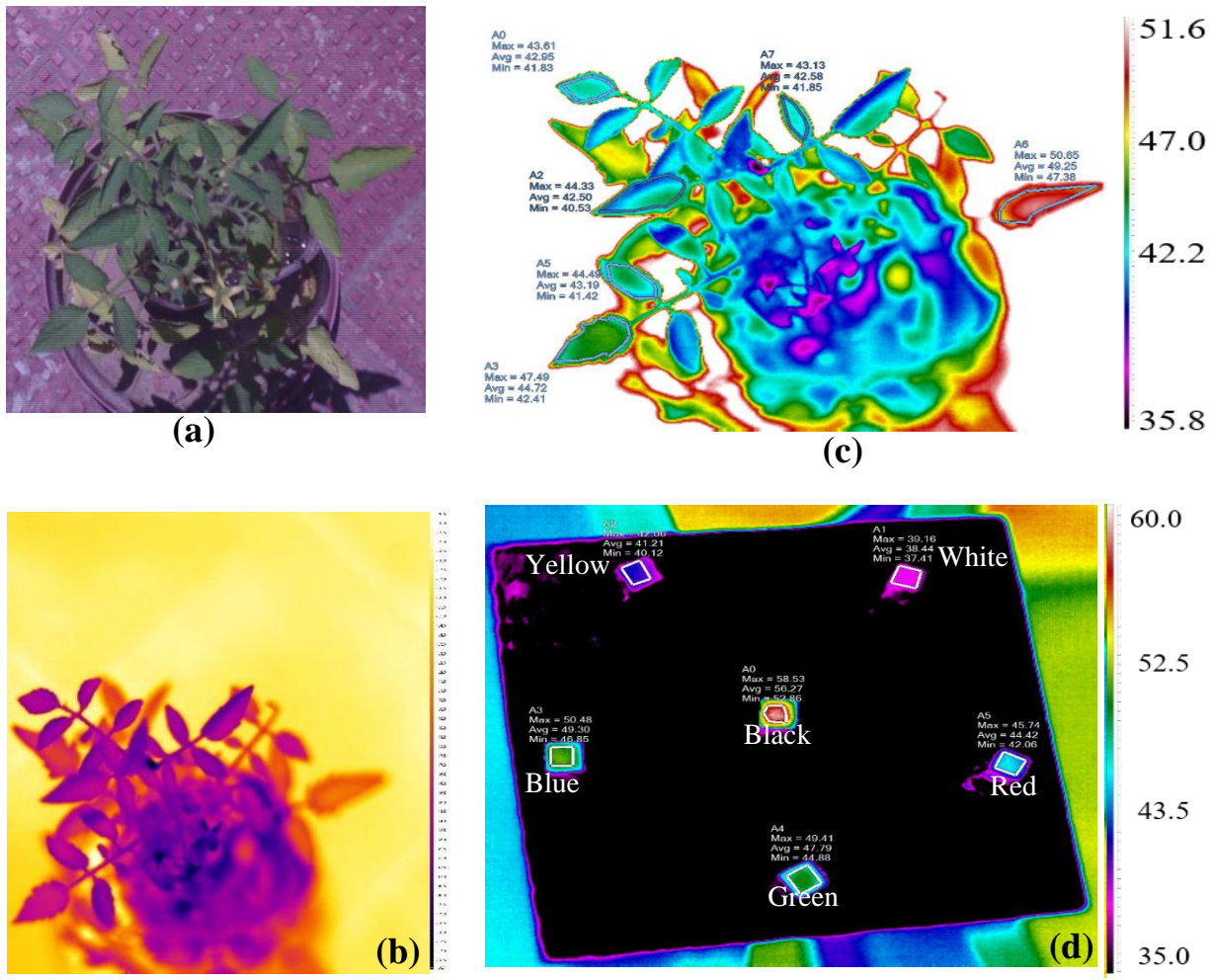

Figure 7 Comparison of a 'dying' leaves temperatures with a set of dry (a) Visible and (b) raw thermal image of the plant, (c) post-processed thermal image of the plant showing individual leaves, and (d) processed thermal image of the dry surfaces. All these images were captured on 3$^{rd}$ June 2023. Interestingly, the rightmost leaf in (c) is at a temperature nearly equal to that of the 'blue dry' surface (see 'd').

Table 5 is the temperature data record of the dying leaves along with both sets of surfaces. The dying leaves' temperature was plotted with respect to each dry and wet surface as seen in Fig. 8a and 8b. The dying leaves data can be seen accumulated closer to the yellow surface and a few to the red dry surface. Temperature values of the white dry surface are now consistently lower when compared to those of the dying leaf temperature data. The maximum temperature values are, as usual, shown by the black dry reference surfaces. As far as the wet reference surfaces are concerned, now, their temperature values are quite low compared to the dying leaves temperature. The least difference in temperature data can be seen between the black wet reference surface data and dying leaf temperature data.

Table 5 Surface temperature of *dying* tomato leaves measured along with different sets of 'dry' and 'wet' surfaces.

| Plant Temperature ($^0C$) | Dry Reference Surfaces Temperature ($^0C$) | | | | | | Wet Reference Surface Temperature ($^0C$) | | | |
|---|---|---|---|---|---|---|---|---|---|---|
| **Dying Leaf** | **White** | **Yellow** | **Red** | **Green** | **Blue** | **Black** | **White** | **Coffee FP** | **Green** | **Black** |
| 38.98 | 37.57 | 41.1 | 41.7 | 45.87 | 47.57 | 53.63 | 30.47 | 30.96 | 31.76 | 32.45 |
| 39.92 | 37.57 | 41.1 | 41.7 | 45.87 | 47.57 | 53.63 | 30.47 | 30.96 | 31.76 | 32.45 |
| 40.08 | 37.57 | 41.1 | 41.7 | 45.87 | 47.57 | 53.63 | 30.47 | 30.96 | 31.76 | 32.45 |
| 41.08 | 38.44 | 41.21 | 44.42 | 47.79 | 49.3 | 56.26 | 31.09 | 31.75 | 33.5 | 37.15 |
| 41.15 | 38.44 | 41.21 | 44.42 | 47.79 | 49.3 | 56.26 | 31.09 | 31.75 | 33.5 | 37.15 |
| 41.79 | 38.44 | 41.21 | 44.42 | 47.79 | 49.3 | 56.26 | 31.09 | 31.75 | 33.5 | 37.15 |
| 42.18 | 38.44 | 41.21 | 44.42 | 47.79 | 49.3 | 56.26 | 31.09 | 31.75 | 33.5 | 37.15 |
| 42.46 | 38.44 | 41.21 | 44.42 | 47.79 | 49.3 | 56.26 | 31.09 | 31.75 | 33.5 | 37.15 |
| 42.62 | 37.57 | 41.1 | 41.7 | 45.87 | 47.57 | 53.63 | 30.47 | 30.96 | 31.76 | 32.45 |
| 43.69 | 39.56 | 40.8 | 43.79 | 45.43 | 46.83 | 50.72 | 29.96 | 31.35 | 33.87 | 35.67 |

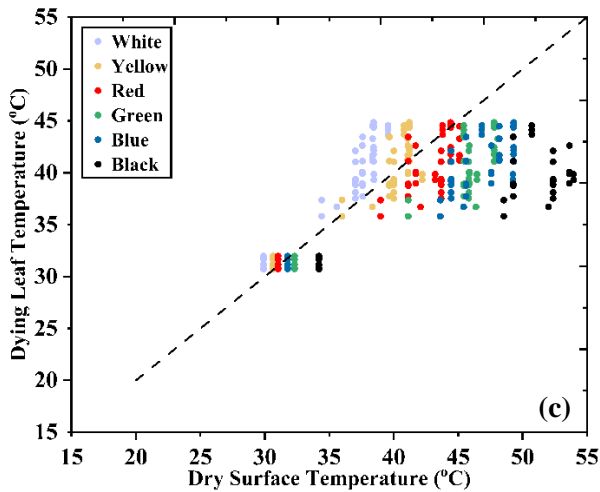
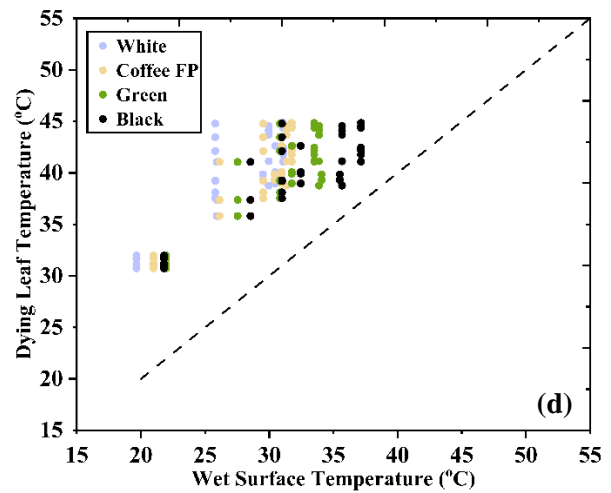

Figure 8 Comparison of *dying leaves* temperature data with (a) dry and (b) wet reference surfaces using $45^0$ slope line. Here, a majority of the leaves temperature are observed to be at temperatures close to that of yellow and red dry surfaces.

### 3.1.3 Dead leaf versus dry and wet reference surfaces

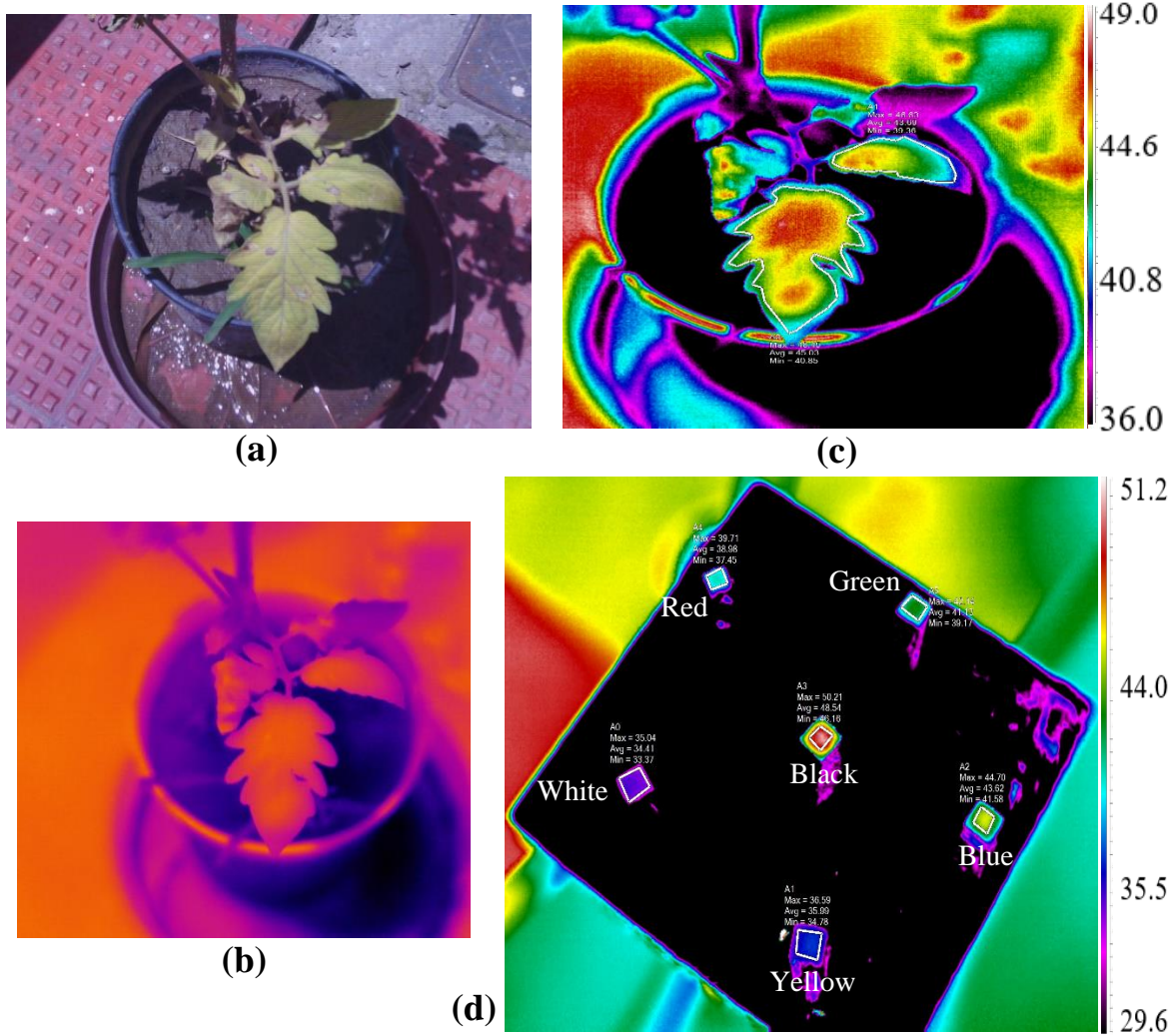

Figure 9 Comparison of *'dead' plant* leaves temperatures with a set of dry (a) Visible & (b) raw thermal image of the plant, (c) post-processed thermal image of the plant showing individual leaves, and (d) processed thermal image of the dry surfaces captured on 8$^{th}$ June 2023. Here, the average temperature of the leaf at the bottom of (c) is close to that of the 'blue dry' surface.

Figure 9a shows the optical image of such a condition where the leaf turned yellowish and appears to be partially dead. After analysing the raw thermal images (Fig. 9b), the temperatures of such leaves were found to be in the range of 45-48°C as seen in Fig. 9c; the temperature of dry surfaces were also recorded at the same instant (Fig. 9d). Here, the dead leaf temperature can be seen coinciding with the green dry surface with temperature difference range varying from 0.01-3.77$^0$C. However, across all the measurements, this temperature difference between blue dry surfaces and dead leaves also falls within this narrower range but mostly below the slope line, compared to above it in case of green. Thus, a range of dry surfaces has been predicted within which all the dead leave's temperature falls, whereas the dead leaf data when plotted with respect to the wet surface temperatures predicted an upward shift as compared to the plots of healthy and dying leaves (Fig. 10b). We can, therefore, conclude that red, green, and blue dry reference surfaces may suitably represent a dead leaf. Black dry surfaces are still very hot and can be excluded from our study hereafter.

Table 6 Surface temperature of dead tomato leaves with different 'Dry' & 'Wet' surfaces.

| Plant Temperature (⁰C) | Dry Reference Surfaces Temperature (⁰C) | | | | | | Wet Reference Surface Temperature (⁰C) | | | |
|---|---|---|---|---|---|---|---|---|---|---|
| Dry/Dead Leaves | White Dry | Yellow | Red | Green | Blue | Black | Wet W | Wet FP | Wet G | Wet BK |
| 41.57 | 34.41 | 35.99 | 38.98 | 41.13 | 43.62 | 48.54 | 25.91 | 26.13 | 27.53 | 28.53 |
| 42.08 | 34.41 | 35.99 | 38.98 | 41.13 | 43.62 | 48.54 | 25.91 | 26.13 | 27.53 | 28.53 |
| 42.69 | 34.41 | 35.99 | 38.98 | 41.13 | 43.62 | 48.54 | 25.91 | 26.13 | 27.53 | 28.53 |
| 43.29 | 34.41 | 35.99 | 38.98 | 41.13 | 43.62 | 48.54 | 25.91 | 26.13 | 27.53 | 28.53 |
| 43.69 | 39.56 | 40.8 | 43.79 | 45.43 | 46.83 | 50.72 | 29.96 | 31.35 | 33.87 | 35.67 |
| 43.69 | 34.41 | 35.99 | 38.98 | 41.13 | 43.62 | 48.54 | 25.91 | 26.13 | 27.53 | 28.53 |
| 44.14 | 39.56 | 40.8 | 43.79 | 45.43 | 46.83 | 50.72 | 29.96 | 31.35 | 33.87 | 35.67 |
| 44.56 | 39.56 | 40.8 | 43.79 | 45.43 | 46.83 | 50.72 | 29.96 | 31.35 | 33.87 | 35.67 |
| 44.9 | 34.41 | 35.99 | 38.98 | 41.13 | 43.62 | 48.54 | 25.91 | 26.13 | 27.53 | 28.53 |
| 45.44 | 39.56 | 40.8 | 43.79 | 45.43 | 46.83 | 50.72 | 29.96 | 31.35 | 33.87 | 35.67 |
| 45.66 | 39.56 | 40.8 | 43.79 | 45.43 | 46.83 | 50.72 | 29.96 | 31.35 | 33.87 | 35.67 |
| 45.66 | 39.56 | 40.8 | 43.79 | 45.43 | 46.83 | 50.72 | 29.96 | 31.35 | 33.87 | 35.67 |
| 47.94 | 39.56 | 40.8 | 43.79 | 45.43 | 46.83 | 50.72 | 29.96 | 31.35 | 33.87 | 35.67 |
| 48.49 | 39.56 | 40.8 | 43.79 | 45.43 | 46.83 | 50.72 | 29.96 | 31.35 | 33.87 | 35.67 |

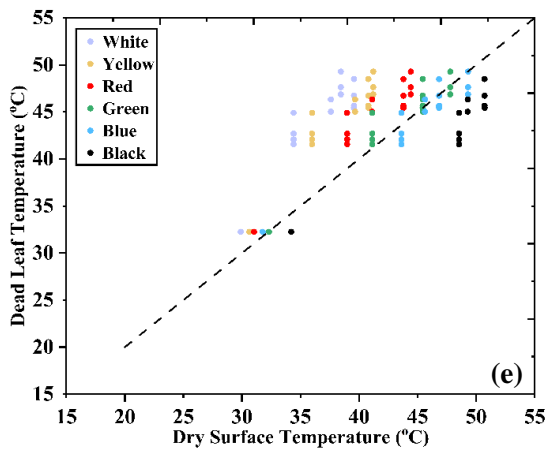 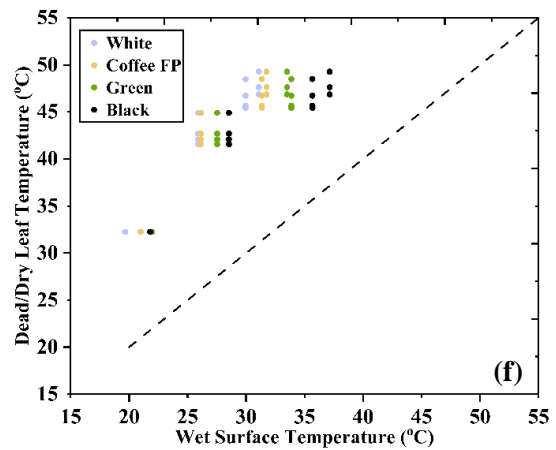

Figure 10 Comparison of *dead leaves* temperature data with (a) dry and (b) wet reference surfaces using 45° slope. Here, the blue and green dry reference surfaces show the least deviation from the slope line, and thus make these surfaces a suitable indicator to represent the dead leaves.

While choosing the best suited reference surfaces for healthy, dying, and dead leaves as discussed earlier, interesting measurement based observations were made which helped to narrow down the range of reference surfaces and to devise a scale to depict the leaves' health based on the dry and wet reference surface temperatures The below bar graphs (Fig. 11a and 11b) represent the scale of temperatures of all the surfaces, i.e., leaves, dry, and wet surfaces on two different days. It was observed that across all the days, the temperature variation of the dry reference surfaces follows a certain trend,

$$White < Yellow < Red < Green < Blue < Black$$

Within this trend, the healthy to dying leaves' temperature most suitably resembled within the range of white to green dry surface temperatures and dying to dead leaves. Thus, the suitability of reference surfaces narrows down to four surfaces i.e., yellow, red, green, and blue.

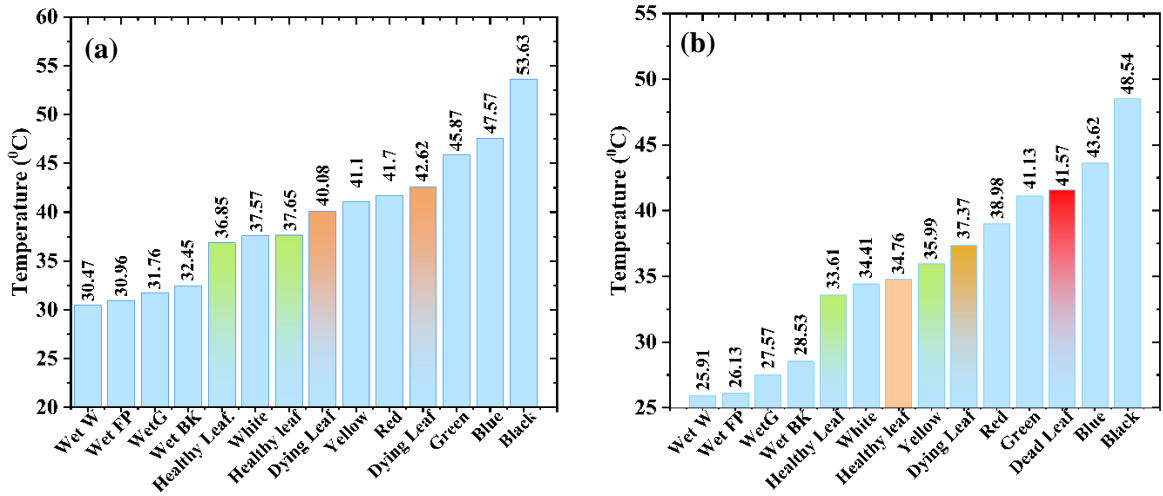

Figure 11 Bar graph representing the leaf temperature data and all the dry and wet surfaces recorded on (a) 5th May 2023 and (b) 3rd June 2023. In both the graphs, the increasing trend in temperature is clearly seen.

### 3.2 Reference surfaces correlating both 'dying' and 'dead' leaves

Figure 12a shows the temperature data of all the dying leaves recorded throughout the experimental period plotted versus the 'yellow' and 'red' dry surfaces. Intriguingly, all the data points cluster around the 45° slope line within a range of ±4°C from the slope line. This clustering suggests that the temperature ranges displayed by the 'yellow' and 'red' dry reference surfaces provide a more accurate indication of dying leaves compared to other surfaces. Furthermore, when the temperature data of dead leaves was plotted against the temperatures profile of red, green, and blue dry surfaces, distinct trends emerged. As a leaf's condition approached complete death, its temperature behaviour converged towards that of the 'green' dry surface and, to a lesser extent, towards the 'blue' dry surface. Figure 11b provides a visual representation of this trend. Notably, larger deviations (~5°C) from the slope line can be observed for the data points of 'Dead leaf' versus 'Red' dry surface. However, a significant portion of the temperature data of the dead leaves closely align with the slope line when plotted against the 'green' dry surface and the 'blue' dry surface temperatures. This suggests that the green dry surface can potentially serve as a more reliable indicator of the health status of dead leaves. It suggests that certain dry surfaces, such as 'yellow' / 'red' for dying leaves and 'green' for dead leaves, exhibit temperature behaviour that can effectively indicate the health condition of the leaves.

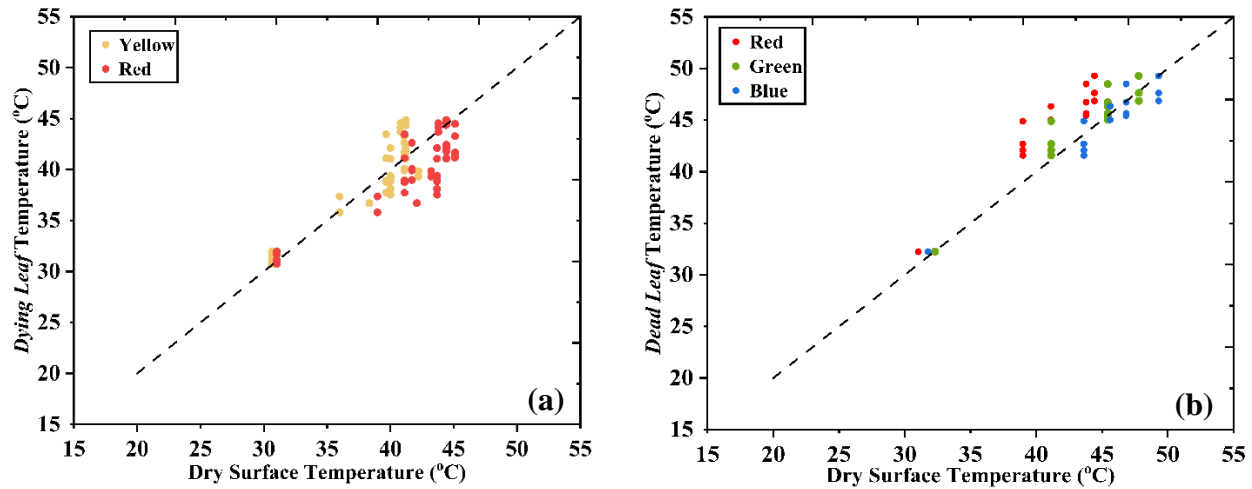

Figure 12 Comparison of (a) dying and (b) dead leaves temperature data with a few chosen dry surfaces such that a majority of data follow the $45^0$C slope line.

### 3.3 Temperature correlations of all leaves with 'Red' and 'Green' dry surfaces

We now look closely at the temperature behaviour of all the leaves condition with respect to the 'red' and 'green' dry surfaces only.

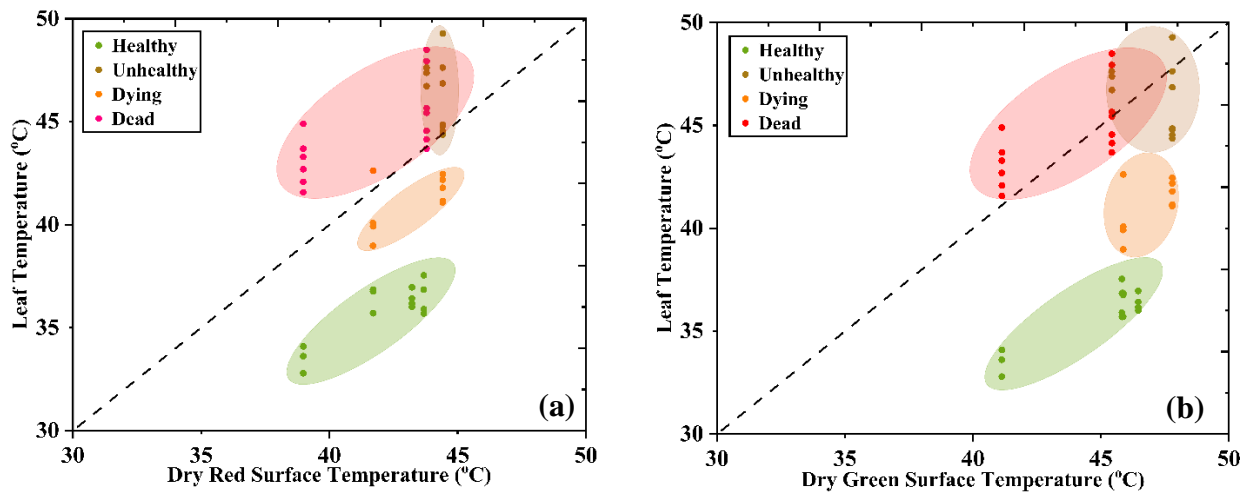

Figure 13 Data showing the temperature correlations of all the leaves versus (a) red and (b) green dry surfaces. The data gives an idea of the range of deviation from the $45^0$C slope line which in turn is an indicator of the water stress in the leaves.

We plotted the temperature data of all the leaves against both the red dry and green dry surfaces. This approach allowed us to pinpoint specific temperature ranges associated with leaves at different stages of health: 'healthy,' 'unhealthy (infected),' 'dying,' and 'dead.' These distinct regions are marked on the temperature plot (Fig. 13a & 13b), offering a visual representation of how leaf temperature correlates with dry surface temperature, particularly in relation to the green and red dry surfaces.

## 3.4 Plant health prediction: Optical Imaging versus IR thermographic measurements

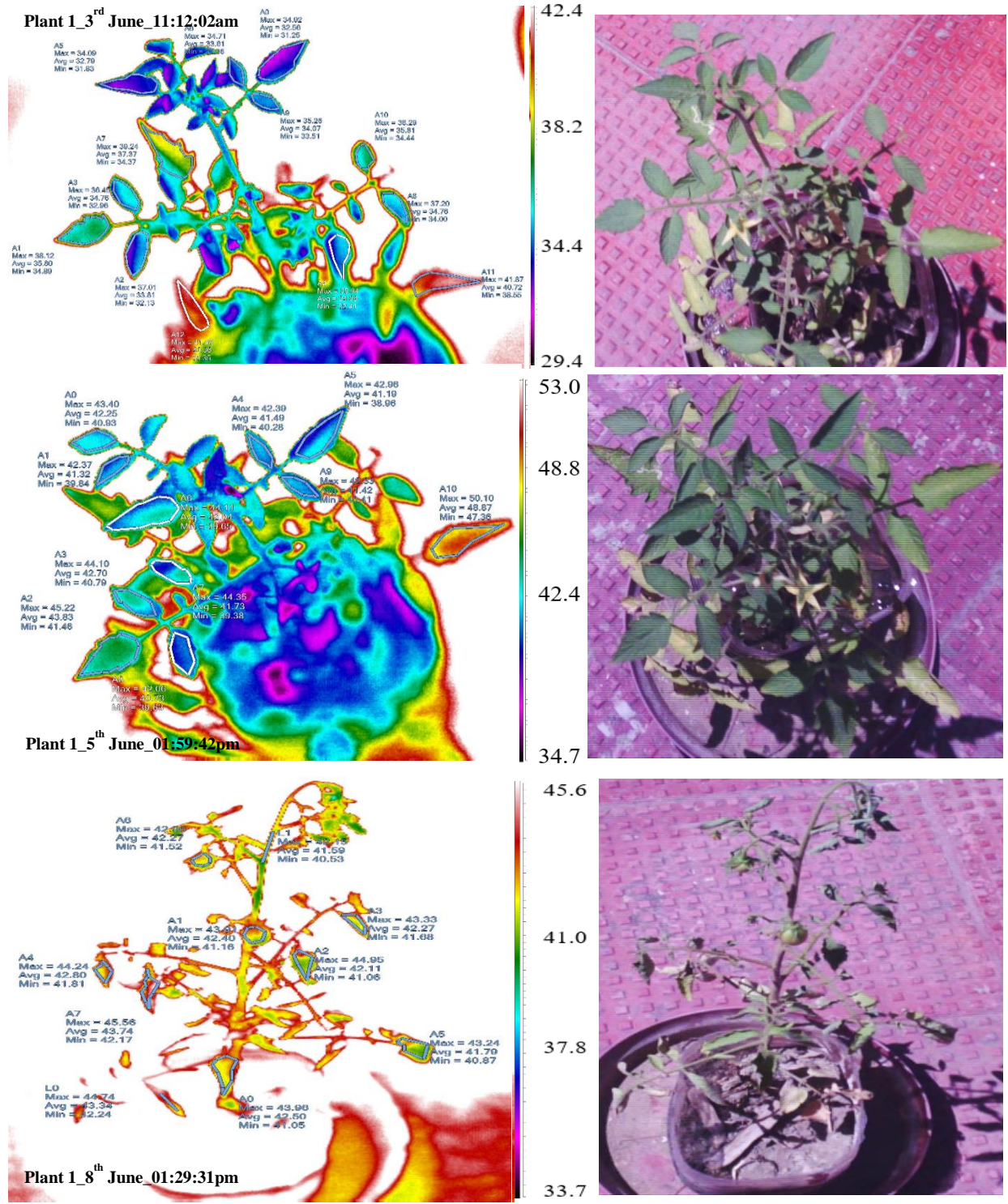

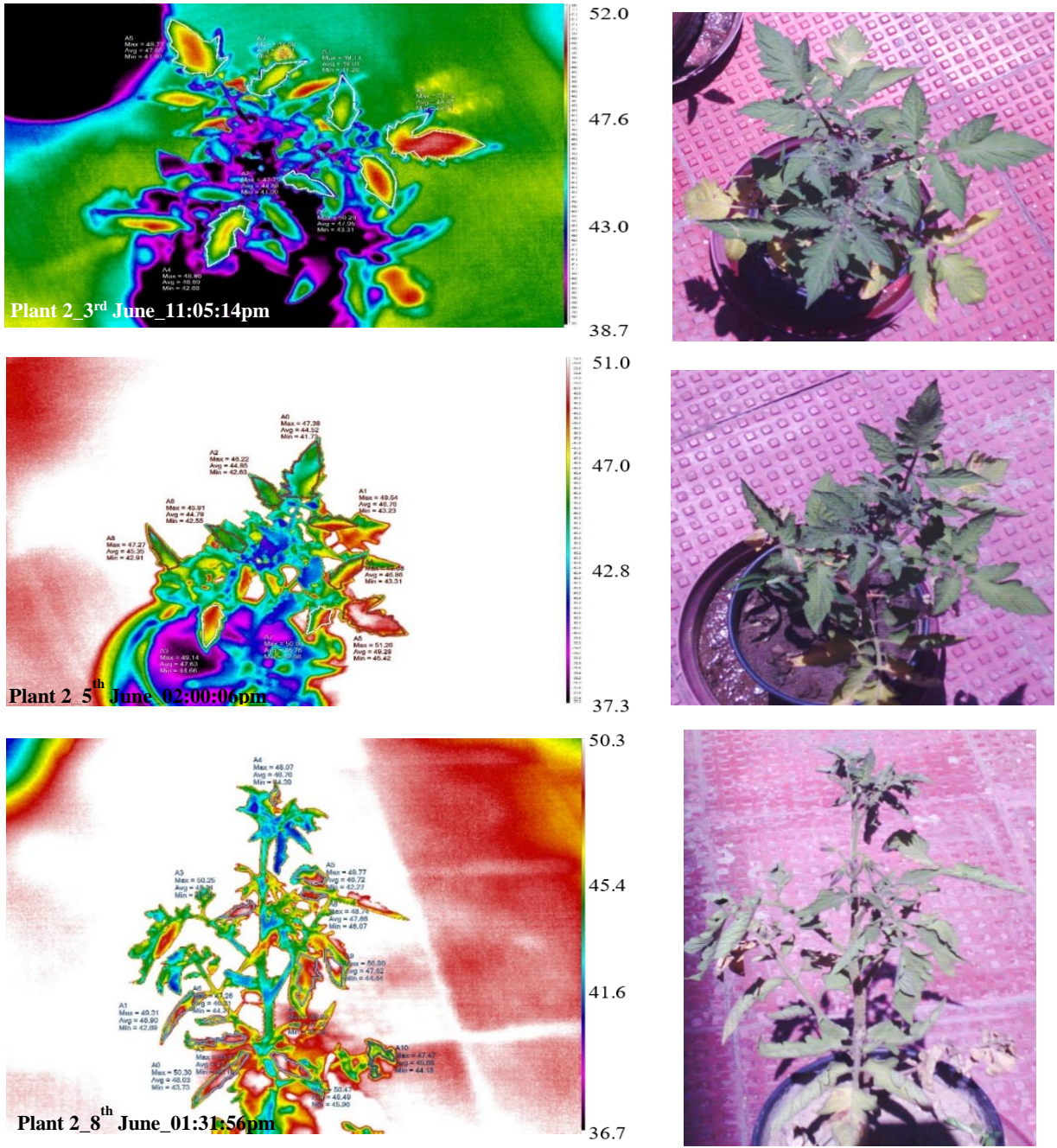

Figure 14 Thermal and optical images of the tomato plant shwoing its journey over a period.

A key difference between the optical and IR imaging is the power of IR imaging in detecting increased water stress in the leaves much sooner than that of the optical symptoms. The increased water stress levels were clearly observed as higher leaf temperatures (see Fig. 14) compared to the fully healthy leaves. IR imaging predicted the unhealthy leaf conditions ~2 days before the same was observed optically (appearance of yellowness of the leaf). Once the optical symptoms were observed, the leaf died (2-3) days afterward. Above, we present visual and IR images for Plant_1 and Plant_2 on June 3rd (refer Fig.14). At this juncture, both plants appeared healthy in the visible spectrum, with most foliage exhibiting a greenish hue. However, closer examination of post-processed IR images revealed higher temperatures in Plant_2, ranging from 44 to 48 $^0$C. This temperature range had been previously correlated with dying leaves. In contrast, the temperatures exhibited by Plant_1 remained within the range characteristic of healthy leaves, approximately 32-37$^0$C. Subsequent to this point, plants were not watered resulting in the increment in their surface temperature but rainfall irrigated them on the June 5$^{th}$ which decreases the plant 1's temperature to 40-43$^0$C. However, Plant_2 displayed no significant drop in temperature even after the rainfall; its temperatures persisted in the range of 44-48$^0$C indicating an attainment of a specific threshold of stress but visually seemed to be healthy. However, interestingly, thermal images consistently displayed elevated temperature on June 3rd and 5th indicating the plants' progression towards under stress and ultimately death/wilted on the 5th day, i.e., June 8$^{th}$. Hence, the temperature data collected throughout these experiments serves as foundational information for the development of a suitable stress index.

## 4. CONCLUSIONS

An extensive investigation into temperature signatures, exhibited by tomato plants and various paper-based dry and wet reference surfaces was conducted. The primary focus was on identifying suitable reference surfaces for assessing plant health based on temperature alone. The study commenced in April, 2023 by capturing thermal images of tomato plants and dry surfaces, with the introduction of wet surfaces in May and June, 2023 to record the potential lower temperatures limit. Experiments were conducted between 11:00 am and 3:00 pm to mimic typical daytime conditions for plant growth, but with no control over the ambient conditions. The findings demonstrated that out of all the dry surfaces, the white dry surface exhibited the lowest temperature, followed by yellow and red surfaces, while the black dry surface displayed the highest temperature. Healthy leaves maintained temperatures within the range of wet black surfaces and white dry surfaces, occasionally registering slightly higher temperatures than the latter on extremely hot days. As plant's health deteriorated, visible changes were evident at a later stage, with some leaves transitioning to a yellowish-green hue. Corresponding thermal images revealed higher temperatures, signifying water stress in the plants. Temperature scales shifted towards higher values, with many leaves reaching 40-44°C, approximately 8-10°C higher than the healthy ones.

The temperature data for dying leaves closely aligned with temperature ranges exhibited by the 'yellow' and 'red' dry reference surfaces. Furthermore, as leaves approached complete demise, their temperature values converged towards the 'green' dry surface and, to a lesser extent, the 'blue' dry surface. These observations suggest that specific dry surfaces, such as 'yellow' and 'red' for dying leaves and 'green' for dead leaves, exhibit temperature behaviours that provide accurate indications of leaf health based on temperature differentials. Overall, these findings offer valuable insights into selecting appropriate reference surfaces for assessing plant health based on temperature variations, with implications for applications in agriculture and crop management. The data is envisioned to be collected in the form of various stress scales and figuring out the most suitable scale is a part of future study.

**ACKNOWLEDGEMENTS**

We thank funding support from SERB (DST, GoI) under the grant SRG/2022/000680.

**CONFLICT OF INTEREST**

The authors declare no conflict of interest with anyone or anybody be it an individual or an organization.

**DATA AVAILABILITY**

Data required to reproduce this research is available at : https://figshare.com/account/home#/projects/177900